# An investigation of subgrid mixing timescale modelling in LES of supersonic turbulent flames

Jian An[a], Jieli Wei[a], Jian Zhang[a], Hua Zhou[a], Zhuyin Ren [a, *]

[a]*Institute for Aero Engine, Tsinghua University, Beijing, 100084, China*

**Abstract**

   The predictive numerical simulation of supersonic turbulent combustion, in which the turbulent intensity is high and the fuel/air mixture is near the flammability limit, remains challenging. An investigation of subgrid mixing timescale modelling in large eddy simulation (LES) of supersonic turbulent flames was carried out in this study. In the Partially Stirred Reactor (PaSR) combustion model, the dissipation-rate-based models estimate the mixing timescale by modeling dissipation rate, which may lead to incorrect estimation of mixing times at different grid resolutions. To solve the existing problem, a gradient-based mixing timescale estimation method was proposed. Subsequently, different approaches of subgrid mixing timescale in PaSR are compared and validated on a supersonic mixing layer, including a dissipation-rate-based model and the proposed gradient-based model. Sensitivity analysis of the model constants and grid resolution was also studied and compared with detailed DNS data. The results highlight the importance of mixing models to correctly handle turbulence/chemistry interactions in supersonic combustion and clearly indicate the proposed model can correctly estimate the subgrid mixing timescale at coarse grids and decline rapidly as it tends to DNS limit.

*Corresponding author: Zhuyin Ren, zhuyinren@tsinghua.edu.cn

# 1. Introduction

Renewed interests in hypersonic flight vehicles for terrestrial travel and space exploration and transport have greatly motivated research efforts toward supersonic combustors [1]. The major challenges in scramjet combustor design are to ensure fast mixing of fuel and oxidizer and high combustion efficiency in a very short residence time, e.g., 1 *ms* [2]. A detailed understanding of the fuel-air mixing and combustion fundamentals in a supersonic flow field is vital to the successful design of supersonic propulsive devices [3].

For supersonic combustion, reliable quantitative measurement is challenging and the simultaneously obtainable quantities of three-dimensional features under supersonic conditions are very limited. Numerical approaches nowadays become the most cost-effective way of investigating scramjet combustion and could provide detailed flow and combustion information. Large eddy simulation (LES), which resolves the large-scale turbulent structures while modelling only the small scales, has achieved significant progress in modeling both non-reactive and reactive supersonic turbulent flows [2].

Various options exist for modeling turbulence-chemistry interaction in supersonic reacting flows, such as the flamelet/progress variable approach [4], the linear eddy model [5], the transported probability density function method [6], the Quasi-Laminar (QL) reaction rate model [7], the Eddy Dissipation Concept (EDC) model [8], and the partially stirred reactor (PaSR) [7]. Recently, Gonzalez-Juez et al. [2] and Fureby [7] reviewed the status of above combustion models in LES. The PaSR combustion model, which can account for finite-rate chemistry and the subtle turbulence-chemistry interaction as well, has proven to be effective for supersonic combustion and has been succesfully used in LES studies of the ONERA [9], DLR [10], and HyShot II scramjet combustors [11]. As discussed in [7], the reacting structures in highly turbulent supersonic flames show a correlation with the dissipative structures of the flow that are typically smaller than the filter width Δ. These flames are in general highly wrinkled and fragmented, and one can assume that they are composed of reacting fine structures with the surroundings being dominated by large-scale structures. Therefore each LES cell can be viewed as a partially stirred reactor containing the homogeneous fine structures, exchanging mass and energy with surroundings. The ratio $\kappa$ of the fine structure volume to the LES cell volume is dependent on to the subgrid chemical and mixing timescales, which require model closure. The proper estimation of the subgrid chemical and mixing timescales remains the primary challenge in the PaSR model.

As far as the scalar mixing timescale is concerned, it is typically assumed to be proportional to the turbulence timescale, i.e., $\tau_m = C_m \tau_{turb}$, with $C_m$ being the model constant. The turbulent timescale $\tau_{turb}$ can be obtained based on the level of turbulence modeling. In the RANS context, different formulations using the Kolmogorov and/or the integral timescales have been proposed [12] such as $\sqrt{\nu/\varepsilon}$ or $k/\varepsilon$, with $k$ being the turbulence kinetic energy, $\varepsilon$ being the turbulence dissipation rate, and $\nu$ being the kinematic viscosity. These dissipation-rate-based RANS formulations have been straightforwardly applied in LES/PaSR for turbulent flames without much in-depth analysis. Various approximations have been proposed to estimate $\varepsilon$, a quantity not directly available and difficult to estimate accurately in LES. This results in case-by-case modifications of $C_m$ (e.g., ranging from 0.001 to 1) to ensure reasonable predictions [13-17]. For example, Afarin et al. investigated the moderate or intense low-oxygen dilution $H_2/CH_4$ flame structure using PaSR with $C_m = 1$ [18], while $C_m = 0.01$ is recommended for the turbulent methane/air diffusion flame [16]. Zhong et al. [17] evaluated the effect of $C_m$ on the simulated high pressure spray flame, showing that the change of $C_m$ from 0.03 to 1 has little effects on the predicted ignition delay time and pressure rise and the flame liftoff length is over-estimated. Moreover, the predictions with the dissipation-rate-based models are highly grid dependent and the manual tuning of $C_m$ over a wide range are often needed for reasonable predictions [43,44]. This poses great challenge for the effective use of PaSR for turbulent flames.

In contrast, many advancements [19-28] have been made on modeling scalar mixing timescales in the context of LES/transported probability density function (TPDF) methods for low-Mach turbulent flames. Instead, the mechanical-to-scalar timescale ratio model is generally modeled based on the gradient of the velocity or passive scalar and does not involve the turbulence dissipation rate. The gradient-based model can properly recover the correct DNS limit of scalar mixing timescales with sufficient fine grid resolution [19-28]. Although the LES/FDF simulations such as the ones for a high-speed piloted premixed jet burner flame [21] and supersonic mixing layers [29] show that the predictions are sensitive to $C_m$ and its optimal value is problem dependent and could vary from 5 to 40, the optimal $C_m$ for a given problem is not grid dependent. To further reduce the uncertainty in the manual specification of $C_m$, dynamic closures [20, 30, 31] have been proposed based on subgrid variance and scalar dissipation rate. More recently, efforts [21, 32] have also been made on modeling the mixing timescale of reactive scalars, such that the enhancement of scalar mixing due to reaction-induced scalar gradients can be properly accounted for.

The ideal properties of the gradient-based model have prompted its preliminary use in LES/PaSR simulations of low-Mach turbulent flames. For example, Parente et al. [33-35] introduced a gradient-based model formulated by sub-filter mixture fraction. The effectiveness of the proposed model with $C_m = 2$ is validated by the detailed experimental data from the Adelaide jet-in-hot co-flow burner flames. Currently, the gradient-based model of subgrid scalar mixing has not been attempted for LES/PaSR simulations of turbulent supersonic flames. Motivated by the above observation, a gradient-based mixing timescale model is formulated for supersonic turbulent flames, with its closure being consistent with LES/PaSR. Its performace is validated with a posteri tests in a supersonic reactive mixing layer with detailed chemistry. In addition, the impact of chemical and mixing timescale closure on flame characteristics are systematically investigated. The following sections describe the methodology, including the PaSR model and the modeling details of the chemical and mixing timescales, , $\tau_c$ and $\tau_m$. Numerical setup and simulation details are given in Section 3. Results and discussion are presented in Section 4. Conclusions are in Section 5.

## 2. Methodology

### 2.1 Overview of LES/PaSR

For compressible gaseous flows involving $N$ chemical species, with spatial filtering, the resulting filtered conservation equations are

$$\frac{\partial \bar{\rho}}{\partial t} + \frac{\partial \bar{\rho}\tilde{u}_j}{\partial x_j} = 0 \tag{1}$$

$$\frac{\partial \bar{\rho}\tilde{u}_i}{\partial t} + \frac{\partial \bar{\rho}\tilde{u}_i\tilde{u}_j}{\partial x_j} = -\frac{\partial \bar{p}}{\partial x_i} + \frac{\partial}{\partial x_j}\left(\bar{\tau}_{ij} - \tau_{ij}^{sgs}\right) \tag{2}$$

$$\frac{\partial \bar{\rho}\tilde{E}}{\partial t} + \frac{\partial}{\partial x_j}\left[(\bar{\rho}\tilde{E} + \bar{p})\tilde{u}_j\right] = \frac{\partial}{\partial x_j}\left[\lambda \frac{\partial \tilde{T}}{\partial x_j} + \tilde{u}_i \bar{\tau}_{ij} - H_j^{sgs} - \sigma_j^{sgs}\right] + \bar{\dot{\omega}}_T \tag{3}$$

$$\frac{\partial \bar{\rho}\tilde{Y}_m}{\partial t} + \frac{\partial \bar{\rho}\tilde{u}_j\tilde{Y}_m}{\partial x_j} = \frac{\partial}{\partial x_j}\left[\bar{\rho}D \frac{\partial \tilde{Y}_m}{\partial x_j} - \tau_{m,j}^{sgs}\right] + \bar{\dot{\omega}}_m (m = 1,\cdots,N) \tag{4}$$

$$\bar{p} = \bar{\rho}R(\tilde{Y})\tilde{T} \tag{5}$$

where, superscript "~" denotes the Favre-filtered parameters, and "-" denotes spatially-filtered parameters, $\bar{\rho}, \tilde{u}_j, \tilde{Y}_m, \bar{p}$ and $\tilde{T}$ are filtered density, velocity, mass fraction of species $m$, pressure and temperature, respectively. In the species transport equation, $D = \mu/\rho S_c$ is the molecular diffusivity, where the laminar dynamic viscosity $\mu = A_s\sqrt{T}/(1 + \frac{T_s}{T})$ is computed with the Sutherland's law and $S_c$ is Schmidt number. $\lambda = \mu C_p/Pr$ is the molecular thermal diffusivity, where $C_p$ is the mixture-specific heat at constant pressure and the Prandtl number $Pr$ is 0.72. $R(\tilde{Y})$ is the mixture gas constant. The filtered viscous stress $\bar{\tau}_{ij}$ is given by

$$\bar{\tau}_{ij} = \mu\left(2\tilde{S}_{ij} - \frac{2}{3}\tilde{S}_{kk}\delta_{ij}\right) \tag{6}$$

where $S_{ij} = \frac{1}{2}\left(\frac{\partial u_j}{\partial x_i} + \frac{\partial u_i}{\partial x_j}\right)$ is the strain rate tensor and $\delta_{ij}$ is the Kronecker operator. The filtered total sensible energy is

$$\tilde{E} = \tilde{e} + \frac{1}{2}\tilde{u}_j\tilde{u}_j + k^{sgs} \tag{7}$$

where $\tilde{e} = \tilde{h}_s - \frac{\bar{p}}{\bar{\rho}}$ is internal energy, $\frac{1}{2}\tilde{u}_j\tilde{u}_j$ is resolved kinetic energy, and $k^{sgs}$ is the sub-grid kinetic energy. The mixture sensible enthalpy $\tilde{h}_s$ is

$$\tilde{h}_s = \sum_{m=1}^{N} \tilde{Y}_m \int_{T_0}^{\tilde{T}} C_{p,m}\, dT \tag{8}$$

The reaction source term in the filtered energy Eq. (4) is

$$\bar{\dot{\omega}}_T = \sum_{m=1}^{N} \bar{\dot{\omega}}_m \Delta h_{f,m}^o \tag{9}$$

where $\Delta h_{f,m}^o$ is the enthalpy of formation for $m$-th species. Different from the energy equation for low Mach flows, the work by the shear $\tilde{u}_j\bar{\tau}_{ij}$, the sub-grid enthalpy flux $H^{sgs}$, the sub-grid scale viscous work $\sigma^{sgs}$ are included in Eq. (3), which are expected to be important for supersonic flows.

The closure of the SGS turbulent stress $\tau_{ij}^{sgs}$ is a major concern in LES. In this study, the SGS stress tensor is modeled using the Boussinesq approximation, i.e.,

$$\tau_{ij}^{sgs} = \bar{\rho}\left(\widetilde{u_i u_j} - \tilde{u}_i\tilde{u}_j\right) = -2\bar{\rho}\nu_t\left(\tilde{S}_{ij} - \frac{1}{3}\tilde{S}_{kk}\delta_{ij}\right) + \frac{2}{3}\bar{\rho}k^{sgs}\delta_{ij} \tag{10}$$

where $\nu_t$ is the SGS turbulent viscosity and $k^{sgs}$ is the SGS turbulent kinetic energy. The one-equation SGS model by Yoshizawa et al. [36] is used to model the SGS turbulent viscosity as

$$\nu_t = C_k \Delta \sqrt{k^{sgs}} \qquad (11)$$

where $\Delta = \sqrt[3]{\Delta_x \Delta_y \Delta_z}$ is the filter width and $C_k$ is a constant. An estimate of the SGS turbulent kinetic energy is obtained by solving a separate transport equation as

$$\frac{\partial(\bar{\rho} k^{sgs})}{\partial t} + \frac{\partial(\bar{\rho} k^{sgs} \tilde{u}_i)}{\partial x_i} = \frac{\partial}{\partial x_i}\left[\bar{\rho}\left(\nu + \frac{\nu_t}{Pr_t}\right)\frac{\partial k^{sgs}}{\partial x_i}\right] + P_k^{sgs} - D_k^{sgs} \qquad (12)$$

where $\nu$ is the laminar viscosity coefficient, $Pr_t$ is the turbulent Prandtl number, the production term $P_k^{sgs} = -\tau_{ij}^{sgs} \partial \tilde{u}_i/\partial x_j$ and the dissipation term $D_k^{sgs} = C_d \bar{\rho} (k^{sgs})^{\frac{3}{2}}/\Delta$. Following the work of Liu et al. [37], the model constants $C_k$ and $C_d$ take the values of 0.02075 and 1.0, respectively.

The sub-grid enthalpy flux $H^{sgs}$ and the sub-grid viscous work $\sigma^{sgs}$ in Eq. (4) are closed as

$$H_j^{sgs} - \sigma_j^{sgs} = -\frac{\mu_t C_p}{Pr_t}\frac{\partial \tilde{T}}{\partial x_j} - (\mu_t + \mu)\frac{\partial k^{sgs}}{\partial x_j} + \tilde{u}_i \tau_{ij}^{sgs} \qquad (13)$$

The sub-grid scalar stresses are approximated using an eddy-diffusivity model, which is written as,

$$\tau_{m,i}^{sgs} = \bar{\rho}\left(\widetilde{u_i Y_m} - \tilde{u}_i \tilde{Y}_m\right) = -\bar{\rho}\widetilde{D}_t \nabla \tilde{Y}_m \qquad (14)$$

where $\widetilde{D}_t$ is the turbulent diffusivity modelled as $\bar{\rho}\widetilde{D}_t = \mu_{t/Sc_t}$. A constant value 0.72 is adopted for both the turbulent Prandtl number $Pr_t$ and the turbulent Schmidt number $Sc_t$ following the work of Liu et al. [37]. These quantities can also be determined by a more accurate dynamic approach as reviewed by Xiao et al. [38]. The former solution is used here just for the sake of simplicity.

Due to the highly nonlinear nature of chemical reaction rates generally governed by the Arrhenius law, the mean reaction rates cannot be directly computed with the mean species mass fractions and temperature without accounting for the contribution of subgrid fluctuations. In the PaSR approach, each computational cell is split into two zones: all reactions occur in one zone, that is, the other zone has no reactions. With the PaSR model, each LES cell is divided into fine structures (denoted by *) in which reactions are assumed to take place, and the nonreactive surroundings (denoted by 0) dominated by large-scale coherent flow structures [7]. This assumption helps to define the reactive volume fraction $\kappa$ which takes a value between 0 and 1, and the reaction rates of the surroundings are considered to be negligible compared to that of the fine structure, then the change in cell composition is determined by,

$$\bar{\dot{\omega}}(\bar{\rho}, \tilde{T}, \tilde{Y}) \approx \kappa \dot{\omega}(\bar{\rho}, \tilde{T}^*, \tilde{Y}^*) + (1-\kappa)\dot{\omega}(\bar{\rho}, \tilde{T}^0, \tilde{Y}^0) \approx \kappa \dot{\omega}(\bar{\rho}, \tilde{T}^*, \tilde{Y}^*) \approx \kappa \dot{\omega}(\bar{\rho}, \tilde{T}, \tilde{Y}) \qquad (15)$$

where $\dot{\omega}(\bar{\rho}, \tilde{T}, \tilde{Y})$ is the net reaction rate computed with the resolved means in LES. The reacting volume fraction $\kappa$ is assumed to be proportional to the ratio of chemical reaction time, $\tau_c$, to the total conversion time in the reactor (i.e. the sum of the mixing and chemical timescale) [7],

$$\kappa = \frac{\tau_c}{\tau_c + \tau_m} \qquad (16)$$

Hence the mixing and chemical timescales are crucial for modeling the mean reaction rates.

*2.3 Subgrid chemical timescale modelling*

Theoretically, the chemical time scales can be computed based on the eigenvalues of the chemical Jacobian $J_{jk} = \frac{\partial \dot{\omega}_j}{\partial Y_k}$, where $\dot{\omega}_j$ is the j-th species net production rate and $Y_k$ is the mass fraction of k *th* species [39]. In practice, to overcome the computational cost associated with the evaluation of the chemical Jacobian and its decomposition, especially for complex systems, many models have been proposed for efficient estimation of the overall characteristic chemical timescale. These models often employ a combination of species concentrations, chemical reaction rates and/or species depletion/destruction rates. In this work, the chemical timescale was defined using the formulation of Golovitchev and Nordin [40] as

$$\tau_c^G = \sum_{r=1}^{n_R} \frac{c_{tot}}{\sum_{n=1}^{N_{S,RHS}} v_{n,r} k_{f,r}} \tag{17}$$

where $c_{\text{tot}}$ is the total species concentration, $k_{f,r}$ is the forward production rate, $v_{n,r}$ is the stoichiometric coefficient of the product species and $n_R$ is the number of reactions, and $N_{S,RHS}$ is number of product species of each reaction. This estimation is closer to the overall characteristic time of chemical reaction system and was widely used and proved to be effective [35, 41].

*2.4 Subgrid mixing timescale modelling*

*2.4.1 Dissipation-rate-based model*

In the RANS context, most mixing timescale models are developed with the turbulent dissipation rate $\varepsilon$ being one of the key quantities. Over the past, these dissipation-rate-based models have been simply extended to LES. For example, Nordin [14] estimated the mixing time scale by using the sub-grid turbulent kinetic energy, $k$, and turbulent dissipation rate, $\varepsilon$, as

$$\tau_m^I = C_m^I \frac{k}{\varepsilon} \tag{18}$$

where $C_m^I$ the model constant with a range of [0.001, 0.3] being reported for turbulent supersonic flames in literature [14]. Afarin et al. [18] proposed a definition that employs the effective viscosity and the turbulent dissipation rate as

$$\tau_m^D = C_m^D \sqrt{\frac{v+v_t^D}{\varepsilon}} \tag{19}$$

where $v$ and $v_t^D$ are laminar kinematic viscosity and eddy viscosity for this model, respectively. In the context of the geometric mean of kolmogorov and integral time scales, Fureby [7] proposed a definition based on the Kolmogorov velocity and the shear timescale, i.e.,

$$\tau_m^M = C_m^M \sqrt{\tau_\Delta \tau_k} \tag{20}$$

where $C_m^M$ is the model constant based on specific flow configurations, $\tau_\Delta = \Delta/v'$ is the sub-grid time scale and $\tau_k = (v/\varepsilon)^{1/2}$ is the Kolmogorov time scale, with $\Delta$ being the computational cell scale and $v' = \sqrt{2k/3}$ being the sub-grid velocity fluctuation.

For all these dissipation-rate-based models, $\varepsilon$ is readily available in RANS, but not in the LES context. It is usually calculated by

$$\varepsilon = \frac{(v')^3}{\Delta} = \frac{C_\varepsilon k\sqrt{k}}{\Delta} \tag{21}$$

where $C_\varepsilon$ is a constant.

### 2.4.2 Gradient-based model

In the context of LES/TPDF, the following mechanical-to-scalar timescale ratio model [25-28] has been widely used for subgrid scalar mixing,

$$\tau_m^G = C_m^G \frac{\Delta^2}{v+v_t^G} \tag{22}$$

where $C_m^G$ is a model constant, the eddy viscosity $v_t^G$ could be estimated with the Vreman's algebraic subgrid model [42] that is given by

$$v_t^G = C_v \sqrt{\frac{B_\beta}{\alpha_{ij}\alpha_{ij}}}$$

$$\alpha_{ij} = \frac{\partial \bar{u}_j}{\partial x_i}, \ \beta_{ij} = \Delta^2 \alpha_{mi}\alpha_{mj}$$

$$B_\beta = \beta_{11}\beta_{22} - \beta_{12}^2 + \beta_{11}\beta_{33} - \beta_{13}^2 + \beta_{22}\beta_{33} - \beta_{23}^2 \tag{23}$$

where $C_v = 2.5 C_s^2$ with the Smagorinsky constant $C_s$ being taken to be 0.1, and $B_\beta$ is the second invariant of a tensor quantity as described in Ref. [42]. This estimation has been successfully applied to low-Mach turbulent flames, including piloted premixed turbulent flames [21, 43] and non-premixed turbulent flames [19]. Meanwhile, the parameter sensitivity of $C_m^G$ in low-Mach flames has been systematically investigated [20, 43]. In the context of LES, the optimal value of $C_m^G$ is between 0.02 and 1.

In this study, the above gradient-based model, Eq. (22), is introduced for LES/PaSR simulations of supersonic turbulent flames, with its performance being assessed against the dissipation-rate-based models (i.e., Eq. 19). This gradient-based model is constructed based on the velocity gradients and does not rely on any other modeled subgrid variables. The predicted subgrid scalar mixing time decays quadratically when approaching the DNS limit [43], i.e., as $\Delta$ approaches 0, $\tau_m^G$ approaches 0 at a rate of $\Delta^2$, and this allows the asymptotically correct filtered reaction rates in the DNS limit. In contrast, as detailed in Section 4.4.1, the dissipation-rate-based model has slower decaying rate $\sqrt{\Delta}$. The impact of different model formulation on the predicted flame charaterstics is systematically quantified in a supersonic turbulent reactive mixing layer.

## 3. Supersonic turbulent reactive mixing layer

*3.1 Flame configuration*

In this work, the canonical supersonic reactive mixing layer is employed for model validation, in which the coupling of compressibility, turbulent mixing, and chemical reactions has profound impact on flame initiation and stabilization. Considering that the experimental data with high temporal/spatial resolution are relatively scarce, DNS data of the supersonic reactive mixing layer is employed as numerical experiment that fully resolves the flow and flame structures. Specifically, the comprehensive database of Ferrer et al.[44] is taken as the benchmark to demonstrate the accuracy of proposed scalar mixing timescale models.

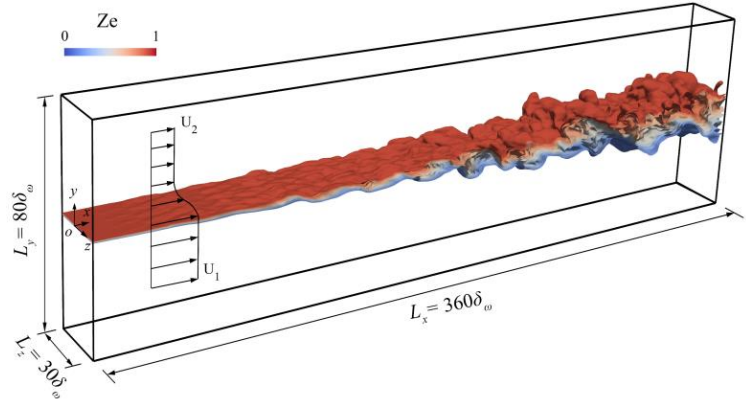

Fig. 1 A scheme of spatially developing mixing layer. *Lx, Ly* and *Lz* are normalized by the initial vorticity thickness $\delta_\omega$ colored by mixture fraction (*Z*).

A schematic view of the supersonic reactive mixing layer is given in Fig. 1, which illustrates the streamwise x, transverse y, spanwise z, directions with the domain lengths $L_x$, $L_y$ and $L_z$ (normalized by the initial vorticity thickness $\delta_\omega$) being given in each direction. The shear layer is formed by two streams, where the bottom stream is supersonic with a mean velocity of $U_1$=1634.0 m/s, while the upper stream has a lower velocity $U_2$=699.1 m/s. The subscripts 1 and 2 refer to the bottom oxidizer stream and the upper fuel stream, respectively. The initial vorticity thickness $\delta_\omega$, the reference length, is 0.988 mm. For such a mixing layer, the compressibility is characterized by the convective Mach number,

$$M_c = \frac{U_1 - U_2}{c_1 + c_2} \quad (24)$$

where *c* is the speed of sound. In the present simulations, the convective Mach number is 0.7. The mean streamwise velocity, species mass fractions and density are initialized with hyperbolic tangent profiles according to the general expression

$$\varphi(y) = \frac{\varphi_1 + \varphi_2}{2} + \frac{\varphi_1 - \varphi_2}{2} tanh\left(\frac{2y}{\delta_\omega}\right) \quad (25)$$

where *φ* denotes any of the flow variables mentioned above.

The other mean velocity components are set to zero while pressure and temperature are set to uniform values. The composition, pressure, temperature, and density in each inlet stream are summarized in Table 1. The detailed hydrogen mechanism of O'Conaire [45], involving 10 chemical species and 21 elementary reactions, is employed to describe the fuel oxidation. At the present flow conditions, the low-concentration intermediate radicals such as O, H, OH in the vitiated air stream are expected to be important for flame initiation and the values in Table 1 are obtained with chemical equilibrium calculation. The specification is consistent with the DNS setting [44], as well as previous analyses of supersonic combustion [46, 47].

Table 1 Flow conditions and stream composition

| Quantity | Fuel stream (top) | Oxidizer stream (bottom) |
| --- | --- | --- |
| Ma | 1.1 | 2.1 |
| Velocity | 669.1m/s | 1634m/s |
| Pressure | 94232.25Pa | 94232.25Pa |
| Temperature | 545.0K | 1475.0K |
| Density | 0.354kg/m3 | 0.203kg/m$^3$ |
| $Y_{H2}$ | 0.05 | 0 |
| $Y_{O2}$ | 0 | 0.278 |
| $Y_{H2O}$ | 0 | 0.17 |
| $Y_H$ | 0 | $5.6\times10^{-7}$ |
| $Y_O$ | 0 | $1.55\times10^{-4}$ |
| $Y_{OH}$ | 0 | $1.83\times10^{-3}$ |
| $Y_{HO2}$ | 0 | $5.1\times10^{-6}$ |
| $Y_{H2O2}$ | 0 | $2.5\times10^{-7}$ |
| $Y_{N2}$ | 0.95 | 0.55 |

*3.2 Simulation settings*

In the present study, an in-house reactive solver developed based on the fully compressible flow solver rhoCentralFoam in OpenFOAM 2.3 is used for the LES simulations. This solver has been extenxively validated for turbulent supersonic flows [48-51]. With this sover, Eqs. 1-4 are discretized according to Gauss theorem with the total variation diminishing (TVD) compatible flux limiter, and the diffusive fluxes are reconstructed using central differencing of the inner derivatives. The discretized equations are integrated in time using the second order Crank-Nicholson scheme.

Dirichlet boundary conditions are imposed at the inlet plane. Periodic boundary conditions are settled in the *z*-direction while non-reflective characteristic boundary conditions are imposed in the *y*-direction as well as at the outlet plane. Finally, the transition of the shear layers is obtained using a low-amplitude, white noise perturbation [44, 52] applied at $(x,y)=(4\delta_\omega,0)$ and superimposed on the transverse and spanwise velocity components. This perturbation has a random phase shift in the spanwise direction and its amplitude is 1% for the non-reactive cases and 0.1% for the reactive cases [44]. Furthermore, a low-amplitude white noise is superimposed on the hyperbolic tangent profile to activate the flow instability. The transverse and spanwise velocity components are perturbed with the same type of disturbance. It is noted that all the low amplitude white noise is added only at transverse locations $|y|\leq4\ \delta_\omega$ and its amplitude is 0.1% [44].

Table 2 Grid parameters.

|  | Lx × Ly × Lz | Nx × Ny × Nz |
|---|---|---|
| DNS by Ferrer [44] | 360×60.6×30 | 1517×325×179 |
| Mesh 1 | 360×80.0×30 | 470×101×47 |
| Mesh 2 | 360×80.0×30 | 564×108×55 |
| Mesh 3 | 360×80.0×30 | 705×136×71 |

Given that the grid filtering size is a parameter in the subgrid mixing timescale models, three sets of nonuniform meshs are set up to investigate the model sensitivity with repsect to grid size. The settings of the computational domain and grids are given in Table 2, where $N_x$, $N_y$ and $N_z$ are the corresponding number of grid points. The $L_x$ and $L_z$ are consistent with the DNS, while the Ly is expanded to $80\delta_{\omega,0}$ to reduce the influence of the upper and lower boundaries on the mixing layer. The coarse, medium and fine mesh has 2.2 million, 3.45 million and 7 million grid celss, respectively. Note that the three sets of nonuniform meshs have a finer grid resolution in the central regions and the grid in the $y$ direction is smoothly stretched toward the boundaries with a maximum stretching factor of approximately 1.1.

In addition to the grid filtering size, the model sensitivity with respect to $C_m^D$ in dissipation-rate-based models and $C_m^G$ in gradient-based models are also investigated. For the dissipation-rate-based model by Afarin et al., a model constant between 0.001 and 1 has been reported for $C_m^D$ in literature [13-15]. For the gradient-based model, a typical range of [0.01, 1] has been reported for $C_m^G$ in low-Mach number turbulent flames and generally good prediction is obtained when $C_m^G$ in the range of [0.01, 0.1]. All the testing cases for grid and model sensitivities are summarized in Table 3. For conciseness, we use M1_D0.01 to denote the case with Mesh 1, the dissipation-rate-based model and $C_m^D = 0.01$. Similarly, M2_G0.1 represent the case with Mesh 2, the gradient-based model and $C_m^G = 0.1$.

Table 3 Mesh and subgrid mixing timecale model settings in LES-PaSR simulations

| Mesh | Subgrid mixing timescale model | Model constant |
|---|---|---|
| Mesh 1/Mesh 2/ Mesh 3 | $\tau_m^D = C_m^D \sqrt{\dfrac{\nu + \nu_t^D}{\varepsilon}}$ | 0.01 |
|  |  | 0.1 |
|  |  | 1 |
|  | $\tau_m^G = C_m^G \dfrac{\Delta^2}{\nu + \nu_t^G}$ | 0.01 |
|  |  | 0.05 |
|  |  | 0.1 |

All the simulations are run for at least five times flow-through time to ensure that the flows reach statistically stationary. Statistics are then collected over about ten flow-through times. These statistical quantities are then averaged in the statistically homogeneous (spanwise) direction.

## 4. Results and discussions

*4.1 Model validation*

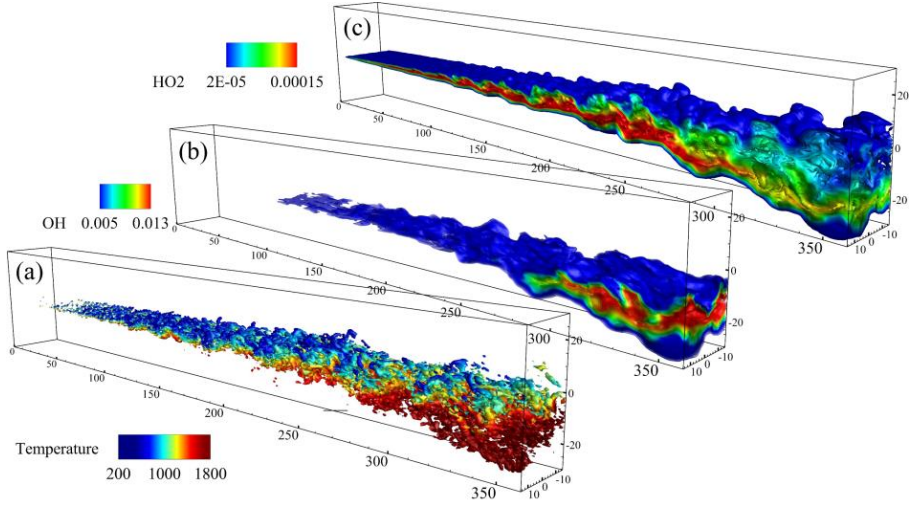

Fig. 2. Iso-surfaces from M3_G0.1 of (a) $\lambda_2$ colored by the local value of temperature, (b) OH and (c) $HO_2$.

To assess the performance of the present solver, the LES-PaSR predictions, obtained with Mesh 3 and the gradient-based model for subgrid mixing timescale, are compared against the DNS results. In addition, LES results with well-stirred reactor combustion model (LES-WSR) and filtered mass density function method (LES-FMDF), conducted by Chen et al. [53] and Guan et al. [29] respectively, are also shown for comparison. Note that the total number of grids in [34, 35] is comparable to that of Mesh 3.

Fig. 2a shows the flow structure visualized by using the isosurface of the $\lambda_2$-criterion. As the shear layer destabilizes, the vortices begin to stretch and break. These complex vortical structures reveal the significance of the three-dimensional obliquely oriented instability of the mixing layer. These complex vortex-structure interactions in general pose challenge for numerical simulations [54]. The consistency in vortical structures with those from DNS proves the applicability of current LES-PaSR for complex flows in supersonic combustion. Considering their senstvitites to the auto-ignition process, radicals OH and $HO_2$ are selected as major flame indicators and their statisics are compared with DNS results to assess the accuracy of the current LES-PaSR accuracy for flame characterstics. As indicated by the instantaneous contours, a pool of $HO_2$ radicals can be readily identified near the injection plane in Fig. 2(c). The early presence of significant amounts of $HO_2$ indicates the beginning of the induction region. The significant growth of radical OH occurs within vortices further downstream. These instantaneous flame characteristics are qualatatvively consistent with those observed in DNS,

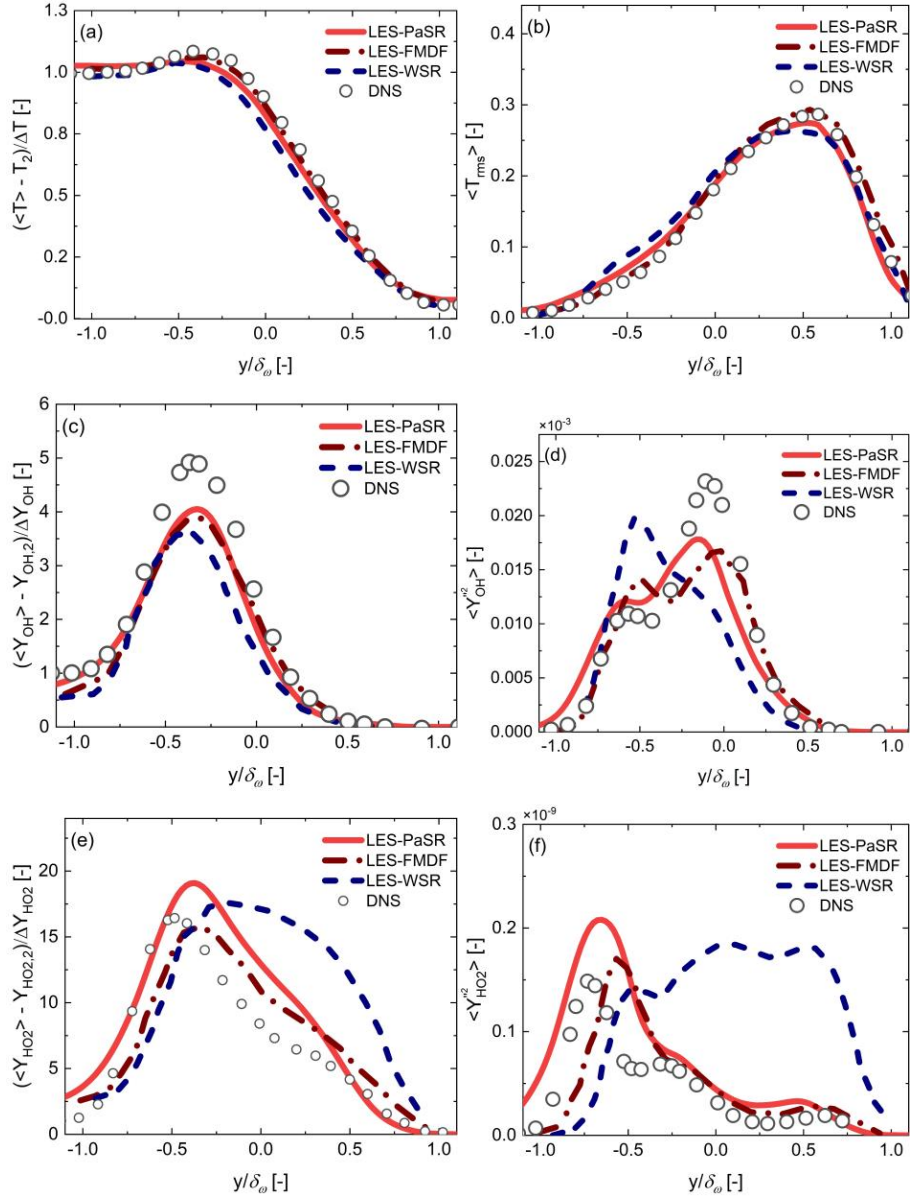

Fig. 3. The mean and variance of temperature, OH and HO$_2$ at $X$=350$\delta_\omega$. The LES-PaSR results are obtained using Mesh 3 and the gradient-based model for subgrid mixing timescale with $\tau_m^G = 0.1$. The data of LES-FMDF, LES-WSR and DNS are reproduced from Refs. [29], [53] and [44], respectively.

Fig.3 further shows the quantantaitve comparsion on the predicted mean and variance of temperature, OH and HO$_2$ from different methods. As shown, compared with DNS, LES-WSR in general underpredicts the combustion intensity, as indicated by its lower predictions in the peak mean temperature, mean OH and HO$_2$ mass fractions at $X$=350$\delta_\omega$. Significant deviations are observed especially for the two important radicals. In Fig. 3(d), the OH fluctuations from DNS show two distinct branches in the mixing layer, i.e., a fuel-lean branch near $y/\delta_\omega$ =

−0.5 and a fuel-rich branch near $y/\delta_\omega$ = 0. LES-WSR presents a single-peak pattern on the oxidant side that is completely different from DNS. For radical HO$_2$, LES-WSR significantly overestimate both the mean and variance for $y/\delta_\omega$ > 0.

In contrast, the current LES-PaSR and LES-FMDF can well reproduce both the mean and variance of tempeature and radical OH and HO$_2$, with comparable accuracy. Both methods successfully reproduce the bimodal/multimodal distributions in the OH and HO$_2$ fluctuations as shown in Figs. 3d and 3f. And the current LES-PaSR approach yields slightly larger peak OH and HO$_2$ compared to LES-FDMF. For this supersonic reacting mixing layer, LES-PaSR achieves comparable predictions as the more sophiscated LES-FMDF.

## 4.2 Effects of model constant $C_m$

Fig. 4 illustrates the sensitivity of the model constant on the time-averaged OH profiles at $X=350\delta_\omega$. For cases with Mesh 1. As shown in Fig. 4(a), for the dissipation-rate-based model, as $C_m^D$ increases from 0.01 to 1, the overall trend of mean OH remains roughly the same, all having significant overestimation on the fuel-lean side. The adjustment of $C_m^D$ is ineffective for improving the predictions since they are in general not senstivie to $C_m^D$ even when it is changed over two order of magnitude. Similar observations on the $C_m^D$ sensitivity are made with Mesh 2 and Mesh 3, as shown in Figs. 4b and 4c. Note that for the dissipation-rate-based model, fine resolution does significantly improve the predictions for the supersonic reactive mixing layer. That is the predictions are grid dependent, which will be further studied in Section 4.3.

In contrast, for the gradient-based model, for cases with Mesh 1, as $C_m^G$ increases from 0.01 to 0.1, the width and peak of the OH distribution show an obvious change, and good agreement with DNS is obtained with $C_m^G = 0.1$. With finer grid resolution Mesh 2 and Mesh 3, although less sensitive, the change in $C_m^G$ still has important impact on the OH profile. That is unlike the dissipation-rate-based model, the predictions from the gradient-based model is sensitive to the model constant $C_m^G$. More importantly, consistent good agreement with DNS can be obtained with $C_m^G = 0.1$ over the three sets of meshes. That is the predictions are less grid dependent, which will be further studied in Section 4.3.

Similar observations on the $C_m^D$ sensitivity are made for OH fluctuation. As shown in Fig. 5a, for the dissipation-rate-based model, on the coarse mesh, the RMS profiles of OH changes insignificantly when $C_m^D$ increases from 0.01 to 1, and the correct bimodal distribution cannot be achieved. Unlike the dissipation-rate-based model, the adjustment of model constant $C_m^G$ has substantial impact on the fluctuation and consistent good agreement with DNS can be obtained with $C_m^G = 0.1$ over the three sets of meshes. Similar observations are made for temperature and other species, which are provided in the Appendix A.

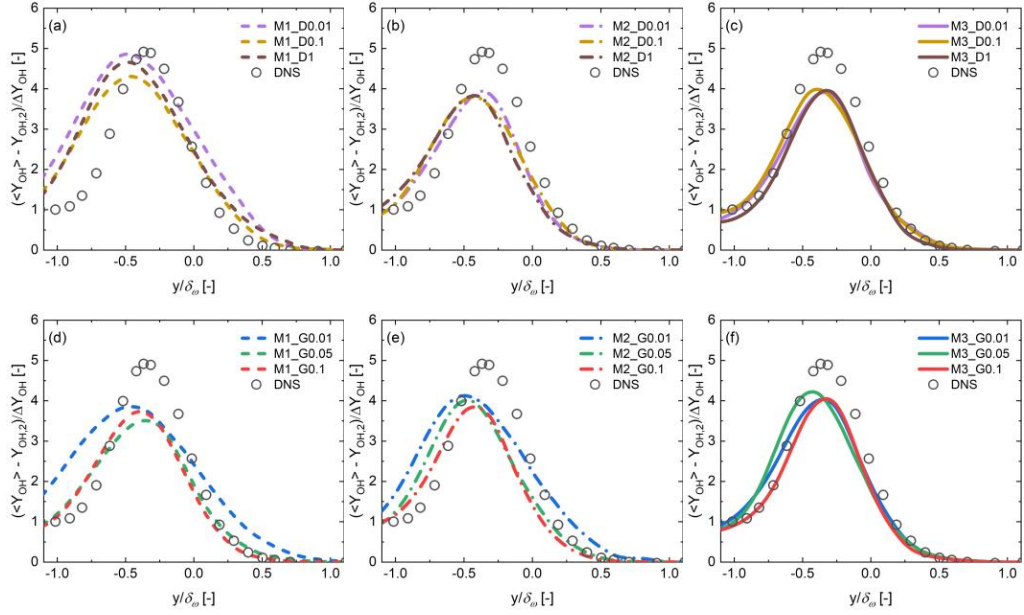

Fig. 4. The profiles of mean OH mass fraction at X=350δ_ω. Top row: the predictions from the dissipation-rate-based model with $C_m^D = 0.01, 0.1, 1$ respectively; bottom row: the predictions from the gradient-based model with $C_m^G = 0.01, 0.05, 0.1$ respectively. Results of DNS are redrawn from Refs. [44].

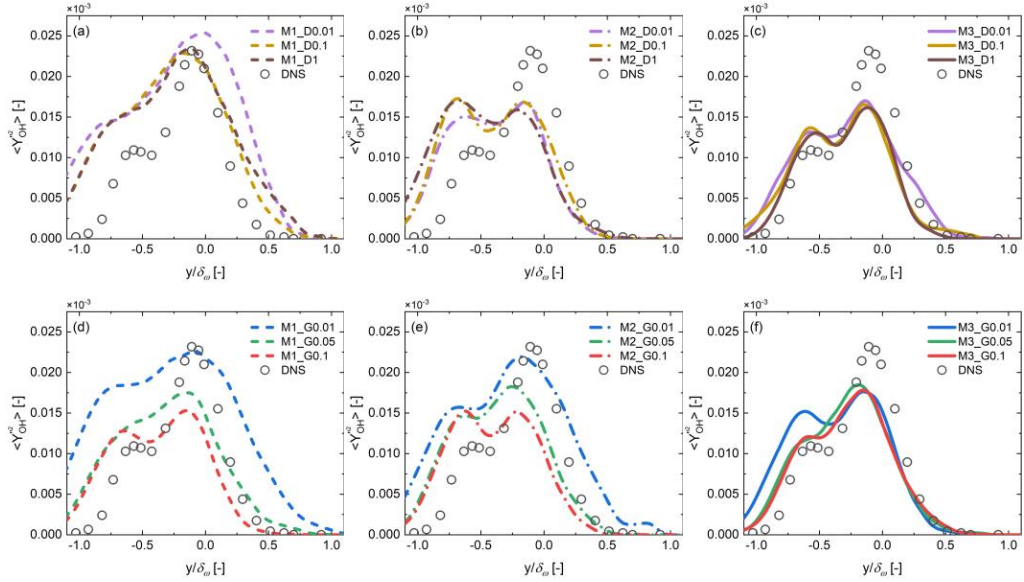

Fig. 5. The profiles of OH rms at *X=350δ_ω*. Top row: the predictions from the dissipation-rate-based model with $C_m^D = 0.01, 0.1, 1$ respectively; bottom row: the predictions from the gradient-based model with $C_m^G = 0.01, 0.05, 0.1$ respectively. Results of DNS are redrawn from Refs. [44].

*4.3 Effects of grid resolution*

Note that the estimation of the subgrid mixing timescale is coupled with grid resolution, Fig. 6 and Fig. 7 illustrate the impact of grid resolution on predictions. As shown, for the dissipation-rate-based model, taking $C_m^D = 1$ as an example, the predictions from the coarse mesh M1 significantly misestimate the intensity of combustion, resulting in wider distribution of OH mean and rms (see Fig. 6c and Fig. 7c). With increasing resolution, better agreement, especially the bimodal distribution of OH variance, can be obtained e.g., with mesh M3. Similar observation are made for $C_m^D = 0.01$ and $C_m^D = 0.1$. That is for the dissipation-rate-based model, the predictions are senstivite to grid resolution. In contrast, for the gradient-based model, the same constant $C_m^G$ yields approximately the same pattern of predictions on different meshes and less grid sensitivity is observed. In particular, with $C_m^G = 0.1$ excellent agreement can be obtained with all three sets of meshes.

The sensitivity study reveals that for the dissipation-rate-based model, the prediction with the same constant $C_m^D$ may vary substantially with grid resolution, and the optimal specification of $C_m^D$ is grid dependent and requires tedious experimentation in practice. In contrast, for the gradient-based model, the optimal constant $C_m^G$ is less grid dependent. Similar observations are made for temperature and other species, which are provided in the Appendix B. Note that the optimal $C_m^G = 0.1$ for this supersonic reactive mixing layer is a typical value used in [21].

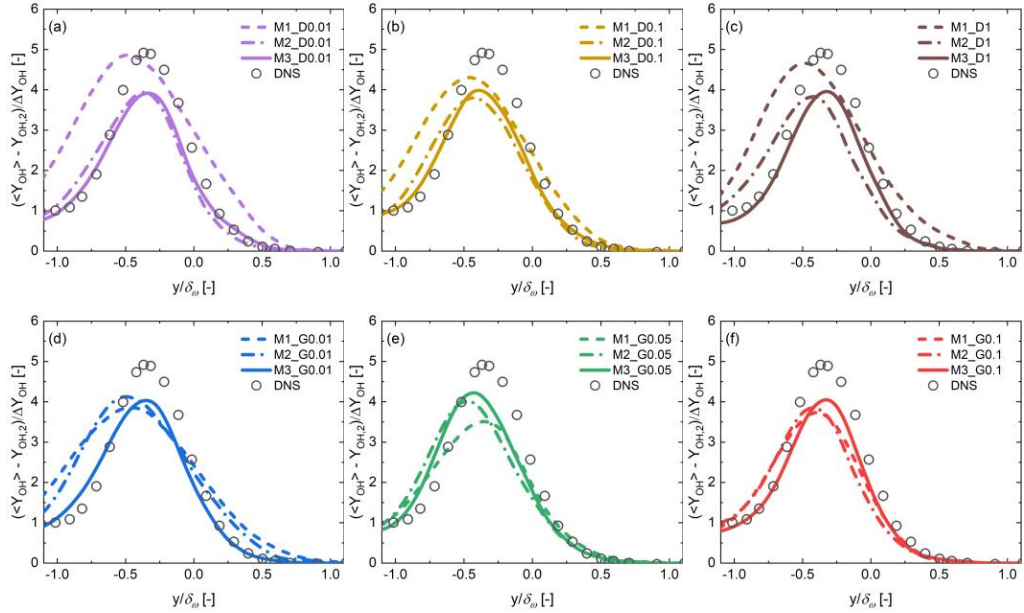

Fig. 6. The profiles of mean OH mass fraction from three meshes at $X=350\delta_\omega$. Top row: the predictions from the dissipation-rate-based model $C_m^D = 0.01$, 0.1, 1, respectively; bottom row: the predictions from the gradient-based model with $C_m^G = 0.01$, 0.05, 0.1, respectively. Results of DNS are redrawn from Refs. [44].

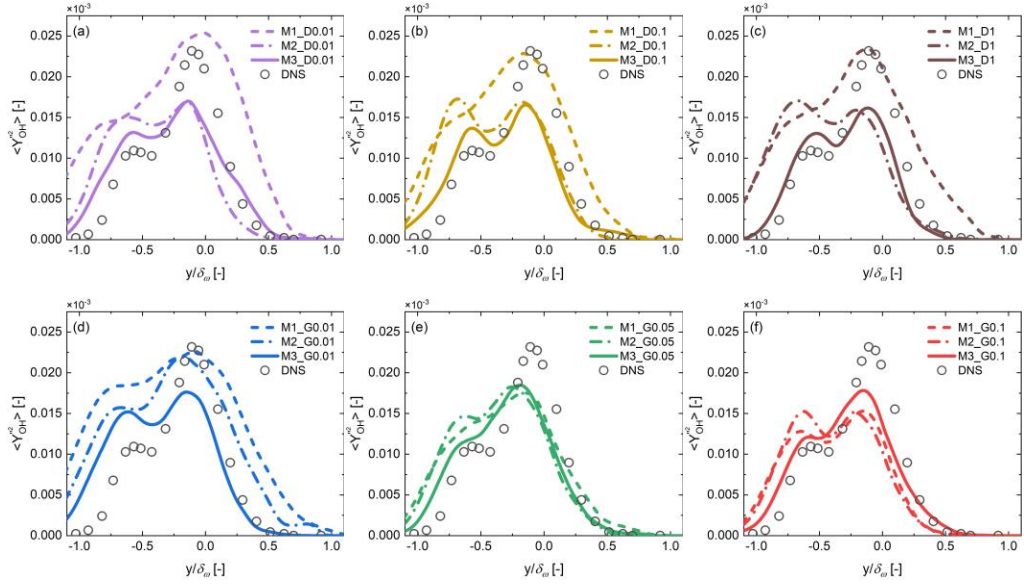

Fig. 7. The profiles of OH rms at $X=350\delta_\omega$ from three meshes at $X=350\delta_\omega$. Top row: the predictions from the dissipation-rate-based model with $C_m^D = 0.01, 0.1, 1$ respectively; bottom row: the predictions from the gradient-based model with $C_m^G = 0.01, 0.05, 0.1$ respectively. Results of DNS are redrawn from Refs. [44].

*4.4 Discussions*

The dissipation-rate-based model and gradient-based model exhibit different prediction sensitivities for model constants and grid resolution. According to Eq. (16) and 错误!未找到引用源。, the closure of turbulence-chemistry interaction is characterized by the variable $\kappa$, which is further determined by the proportionality between subgrid chemical timescale $\tau_c$ and subgird mixing timescale $\tau_m$. In the following, the timescales $\tau_c$ and $\tau_m$ are extracted from the simulations for further analysis.

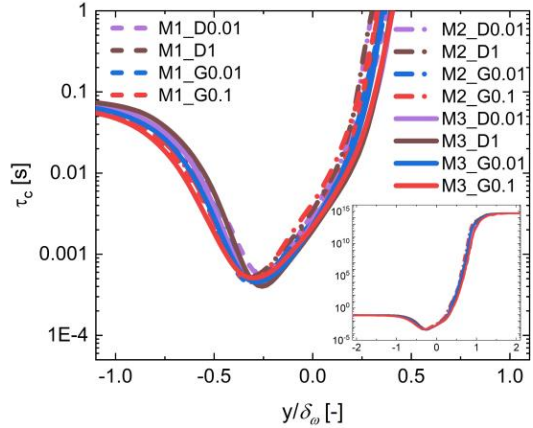

Fig. 8. The profiles of $\tau_c$ from different cases at $X=350\delta_\omega$.

For the subgrid chemical timescale $\tau_c$, it is an overall characteristic time based on the chemical reaction model and should not depend on the grid resolution and model constants. This is confirmed in Fig. 8, which show the extracted $\tau_c$ at the x=350$\delta_\omega$ from all the simulation cases. The characteristic chemical timescales are all in the order of 0.1ms in the region with intense reaction. This is consistent with the characteritic autoignition delay time of hydrogen mixture at the given inflow condition.

*4.4.1 Subgrid mixing timescale*

The difference in subgrid mixing timescale $\tau_m$ is examined. According to Eq.(19) and (21), $\tau_m$ in the dissipation-rate-based model is mainly controlled by the model constant $C_m^D$, grid filter width $\Delta$ and kinetic energy $k$, while the gradient-based model is determined by the model constant $C_m^G$, grid filter width $\Delta$ and algebraic transformation of velocity gradient.

For the influence of the model constant $C_m$, Fig. 9 shows the variation of $\tau_m$ using the same mesh M3. The mixing time scale increases proportionally with $C_m$ as expected. More interestingly, the shape of the profiles are quite different with one having a U-shaped distribution and the other having a ramp-shaped distribution. The difference is correlated with the different flow quantities used in the model formulation. For the dissipation-rate-based model, the shape is governed by the kinetic energy $k$ that is large in the center, but low on the edge of the mixing layer. The mixing timescale varies several orders of magnitude from the edge to the center. In contrast, for the gradient-based model, the shape is rather determined by the mean velocity gradient across the mixing layer and it shows a slight increase from *y/δ$_\omega$* = -1 to 1 corresponding to Ma varying from 2.1 to 1.1. In the bulk region with intense reaction, with the same value of model constant $C_m$, the mixing time from the dissipation-rate-based model is about half of the one from the gradient-based mixing time. That is to achive the same characteristic mixing time in the bulk region with intense reaction, $C_m^D$ should be as twice large as $C_m^G$.

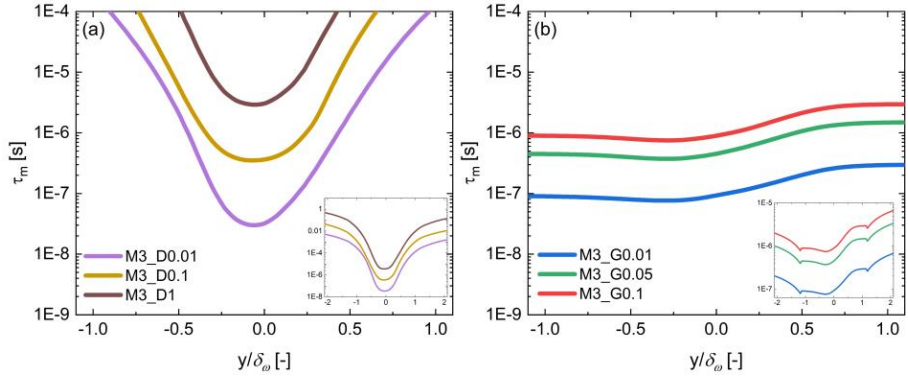

Fig. 9. The variation of $\tau_m$ with model constant $C_m$ from the dissipation-rate-based model Eq. 19 (left) and the gradient-based model Eq. 21 (right) using the same mesh M3

For the dependence on grid filter width $\Delta$, with $\tau_m^D = C_m^D \sqrt{(\nu + \nu_t^D)/\varepsilon}$ (Eq. (19)), $\nu_t^D = C_k \sqrt{k} \Delta$ (Eq. (11)), $\varepsilon = C_\varepsilon k \sqrt{k}/\Delta$(Eq. (21)), one has

$$\tau_m^D \approx C_m^D \sqrt{P_1 \Delta + P_2 \Delta^2} \qquad (26)$$

where $P_1 = \nu/C_k k\sqrt{k}$ and $P_2 = C_\varepsilon/C_k k$. This suggests that

$$\tau_m^D \approx O(\sqrt{\Delta}). \tag{27}$$

Similarly, for the gradient-based model, with Eq. (22) and Eq. (23), one has

$$\tau_m^G \approx \frac{C_m^G}{\frac{M_1}{\Delta^2}+M_2}, \tag{28}$$

where $M_1$ and $M_2$ can be assumed to be independent of $\Delta$. This suggests that

$$\tau_m^G \approx O(\Delta^2). \tag{29}$$

The above dependence on $\Delta$ is confirmed Fig. 10, where as $\Delta$ decreases from $6\times 10^{-5}$ m (M1) to $4\times 10^{-5}$ m (M3), the characteristic $\tau_m^D$ decreases from $3.5\times 10^{-6}$s to $2.9\times 10^{-6}$s and $\tau_m^G$ decreases from $1.6\times 10^{-6}$s to $7.4\times 10^{-7}$s, resulting in a raito of 1.21 times and 2.16 times, respectively. The ratios are consistent with the estimated values of 1.22 (Eq. (27)) and 2.25 (Eq. (29)). If $\Delta$ tends to the DNS limit (about $4\times 10^{-6}$ m), then $\tau_m^D$ and $\tau_m^G$ decrease to $9.2\times 10^{-7}$s and $7.4\times 10^{-9}$s, respectively. The rapid decrease of subgrid mixing timescale ensures that as the grid resolution tends to DNS limit, the gradient-based model can give the asymptotically correct filtered reaction rate i.e., $\tau_m^G \to 0, \kappa \to 1, \bar{\dot{\omega}} = \dot{\omega}$.

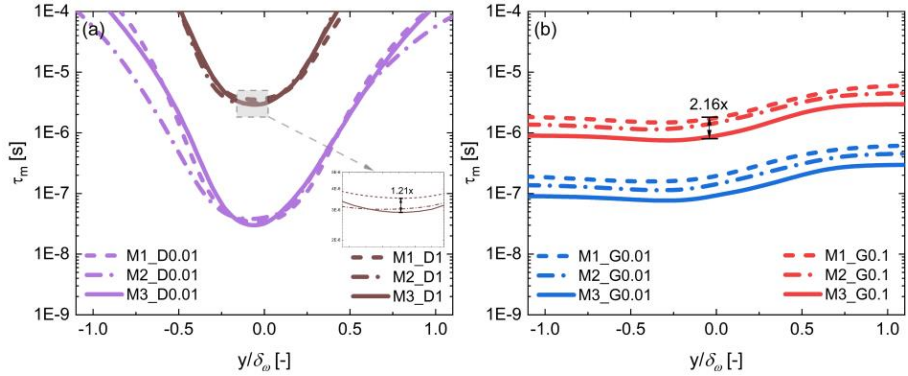

Fig. 10. The variation of $\tau_m$ with grid resolution from the dissipation-rate-based model (left) and the gradient-based model (right).

*4.4.2 Volume fraction κ*

For the volume fraction $\kappa$, its shape is affected by the timescale ratio $\tau_m/\tau_c$. Fig. 11 shows scattered plots of instantaneous temperature and $\kappa$ versus mixture fraction at different cross-sections from the dissipation-rate-based model with $C_m^D = 1$ and the gradient-based model with $C_m^G = 0.1$.

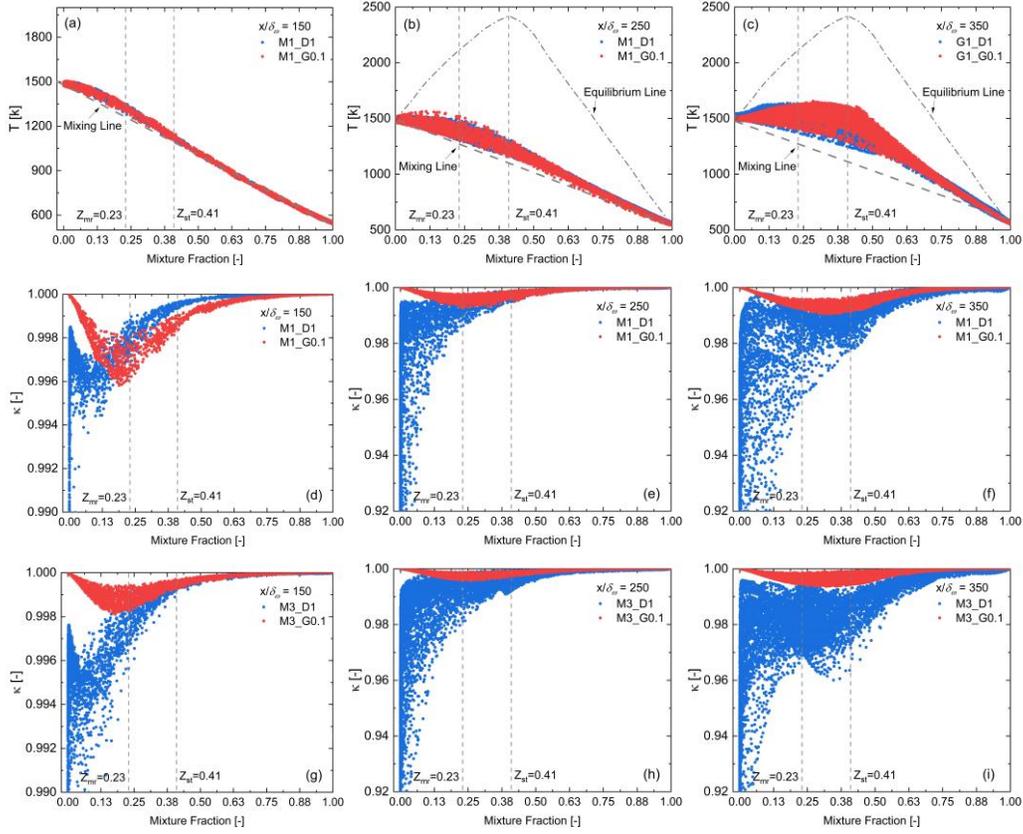

Fig. 11. Scattered plots of instantaneous temperature and $\kappa$ versus mixture fraction ($Z$) at different cross-sections from the dissipation-rate-based model with $C_m^D = 1$ and the gradient-based model with $C_m^G = 0.1$.

The scattered plots of $\kappa$ versus mixture fraction ($Z$) at different cross-sections from the dissipation-rate-based model with $C_m^D = 1$ and the gradient-based model with $C_m^G = 0.1$ are given in Fig. 11d~i. As for the gradient-based model (red scatters in Fig. 11d~i), the $\kappa$ values exhibit a clearly and logically V-shaped distribution. At Z=0 or 1 side, the $\tau_m \gg \tau_c$ (See Fig.8 and 9b), thus the value of $\kappa \to 1$. As the combustion progresses, the position of the valley value moves from $Z_{mr}$ to $Z_{st}$ and its value increases, indicating that the influence of turbulence is gradually weakened, which is consistent with the physical significance. Meanwhile, comparison of the results for M1_G0.1 and M3_G0.1 is full agreement with the above conclusions on the variation of mixing time with grid resolution (see Eq. (29)), i.e., the $\kappa$ obtained by the gradient-based model approaches 1 as $\triangle$ decreases, e.g. in Fig. 11d and g, the minimum value increases from 0.996 to 0.998. In contrast, the $\kappa$ values' distribution pattern of the dissipation-rate-based model is not that irregular (blue scatters in Fig. 11d~i). At $Z=0$, $\tau_c \lesssim \tau_m$, then $\kappa \to 0$, while on the other side, $\tau_c \gg \tau_m$ then $\kappa \to 1$ (See Fig.8, 9a and 10). Along the flow direction, the scatter also gradually moves towards $Z_{st}$, and the values of $\kappa$ are mostly smaller than that of the result of gradient-based model. The difference in the shape of $\kappa$ is correlated with the different flow quantities used in the model formulation, resulting in different estimates of $\tau_m$ (as shown in Fig. 10 and discussed in Section 4.4.1), which ultimately results in the shape of $\kappa$.

Overall, the dissipation-rate-based model leads to a small value of $\kappa$ due to the incorrect estimation of the $\tau_m$ morphology and degradation, i.e., it overestimates the inhibition of chemical reactions by turbulence, which eventually leads to a downstream shift of combustion under a coarse grid, exhibiting a lower temperature distribution and incorrect species distribution. In contrast, the gradient-based model exhibits a logical distribution shape, while being able to correctly estimate $\tau_m$ and $\kappa$ as the grid decreases, allowing predictions on different grids with the optimal constants $\tau_m^G = 0.1$ to be consistent with the DNS results.

## 5. Conclusions

A gradient-based mixing timescale model has been developed for LES/PaSR simulations of supersonic turbulent flames. The scalar mixing timescale is modeled as a function of the local gradient of velocity, thus avoiding the need to model the dissipation rate in LES, and producing no spurious scalar variance when approaching the DNS limit.

The model performance has been investigated in the supersonic mixing layer, together with a comparison with the conventional dissipation-rate-based model. With the gradient-based model, for this supersonic reacting mixing layer, LES-PaSR achieves comparable predictions as the more sophiscated LES-FMDF. Results also show that the chemical time scales with different model constant and grid resolution are similar. The mixing timescale dominates the fine-structure volume fraction and thereafter has a considerable impact on the predicted flame structure.

The parametric study shows that for the dissipation-rate-based model, at coarse resolution, the adjustment of $C_m^D$ is not effective for improving the predictions since it cannot avoid the inherent deficiency in the distribution of subgrid mixing time scale. Only when the resolution is sufficient fine, one can obtain good agreements with DNS data for the supersonic reactive mixing layer, which limits the effective use of PaSR for supersonic turbulent flames.

In contrast, the gradient-based estimation with $C_m^G = 0.1$ on all tested grids achieves excellent agreement, because it can correctly estimate the $\tau_m$ under different grids and does not produce pseudo $\tau_m$ estimates when the grid resolution tends to DNS. In addition, the optimal range of model parameter $C_m^G$, e.g., [0.01,0.1] is consistent with the findings for low-Mach number turbulent flames. In conclusion, with a suitable $C_m^G$, the proposed model can correctly estimate the subgrid mixing timescale at coarse grids and decline rapidly as it tends to DNS limit, and finally model the turbulence-chemistry interaction reasonably. Future work will focus on developing dynamic closures to reduce the uncertainty due to the manual specification of model parameter $C_m^G$ in supersonic turbulent flames.

## Acknowledgements

The work is supported by National Natural Science Foundation of China 52106165 and 52025062.